\documentclass[
a4paper,
amsmath,amssymb,
unpublished,
preprint,
aps,
accepted=2023-12-11,
]{quantumarticle}

\pdfoutput=1
\usepackage[utf8]{inputenc}
\usepackage[english]{babel}
\usepackage[T1]{fontenc}
\usepackage{amsmath}
\usepackage{hyperref}
\usepackage[numbers,sort&compress]{natbib}
\usepackage{ulem}

\usepackage{xcolor, soul} % color and strike text

\usepackage{graphicx}% Include figure files
\usepackage{dcolumn}% Align table columns on decimal point
\usepackage{bm}% bold math

\newcommand{\fig}[1]{Fig.~\ref{fig:#1}}
\newcommand{\eq}[1]{Eq.~\ref{eq:#1}}
\newcommand{\tab}[1]{Table~\ref{tab:#1}}

\begin{document}
\title{Synergy between 
deep neural networks and %\\
the variational Monte Carlo method
for small $^4$He$_N$ clusters}

\author{William Freitas}
 \orcid{https://orcid.org/0000-0002-8020-2117}
 \email{wfsilva@ifi.unicamp.br}
\author{S.~A.~Vitiello}
 \orcid{https://orcid.org/0000-0002-8216-6033}
 \affiliation{Instituto de F\'{i}sica Gleb Wataghin\\
University of Campinas - UNICAMP\\
13083-859 Campinas - SP, Brazil}
%\date{\today}
\maketitle

%\newpage

\begin{abstract}
{
We introduce a neural network-based approach for modeling wave functions that satisfy Bose-Einstein statistics. Applying this model to small $^4$He$_N$ clusters (with $N$ ranging from 2 to 14 atoms), we accurately predict ground state energies, pair density functions, and two-body contact parameters $C^{(N)}_2$ related to weak unitarity. The results obtained via the variational Monte Carlo method exhibit remarkable agreement with previous studies using the diffusion Monte Carlo method, which is considered exact within its statistical uncertainties. This indicates the effectiveness of our neural network approach for investigating many-body systems governed by Bose-Einstein statistics.}
\end{abstract}

\section*{}

The field of artificial intelligence (AI) has seen significant advancements in recent years, making it a promising tool for addressing complex problems in quantum many-body physics \cite{yan20,pfa20,her20,kes21,pes22}. Studies have demonstrated the effectiveness of AI in a wide range of applications \cite{kre22}, leading to increasing interest in exploring its potential for understanding nature as a whole. This paper builds upon this progress by proposing a neural network-based approach for modeling a wave function that satisfies Bose-Einstein statistics and applying it to the study of small clusters of $^4$He$_N$ with $N=\{2,\ldots 14\}$ atoms using the variational Monte Carlo (VMC) method.

Exact solutions of the Schrödinger equation are known for only a limited number of systems. In this context, quantum Monte Carlo methods have become a powerful tool for understanding quantum many-body systems. In recent years, machine learning, specifically neural networks, has gained traction as a method for studying fermionic systems, after the seminal work of Carleo and Troyer  \cite{car17sc}.
Although there have been some efforts \cite{rug18,sai18}, the application of machine learning to the analysis of bosonic systems has generally received less attention. The present work aims to address this gap. Representing trial functions with neural networks and training them through an unsupervised VMC method yields results comparable to those obtained through the diffusion Monte Carlo (DMC) method, highlighting the potential of machine learning for studying Bose-Einstein systems. 
The diffusion Monte Carlo method, a projective method, requires configurations drawn from a trial function that is not orthogonal to the wavefunction of the system's ground state. Even when the available trial function is of low quality, this method can still yield very good results for the ground state  energy of bosonic systems. However, in such cases, estimates of other non-commuting properties with the system's Hamiltonian may be biased due to the need for extrapolation in the estimation process. An alternative approach is to use a neural network-based representation of the ground state trial function, which can provide excellent energies and most likely better estimates of many properties in agreement with experimental results. This study of $^4$He$_N$ clusters is not only a proof of concept, but also a topic that continues to be of recent interest  \cite{yat22,toe22,rec22,spr21,ode21,baz20,kie20,baz19,kie17,car17}.

\section*{Methods}

\begin{figure}
    \centering
\includegraphics[width=.4\textwidth]{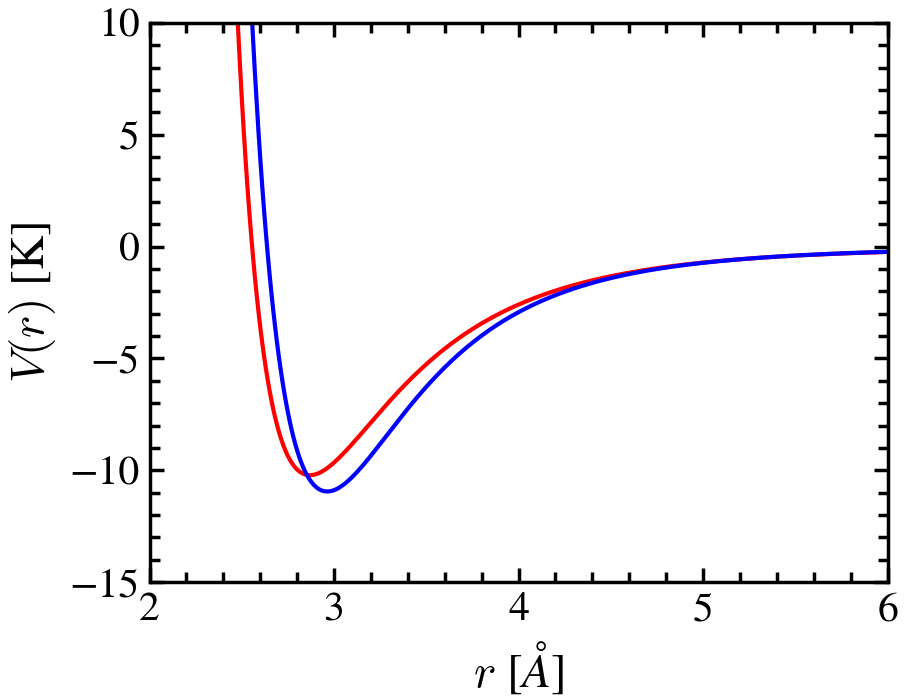}
    \caption{{Potential energy $V$ as a function of the pair separation $r$ for the interatomic potentials Lennard-Jones (red) and HFD-He (blue).} }
    \label{fig:fig01}
\end{figure}

The stationary states of small $^4$He$_N$ clusters can be modeled by assuming that they are described by the Hamiltonian
\begin{equation}\label{eq:hamil}
\mathcal H =
\frac {-\hbar^2}{2m}
\sum_i^N
\bm\nabla_i^2 +
\sum_{i < j}^N
V(r_{ij}),
\end{equation}
where the kinetic energy term  depends on the atomic mass $m$, and $V$, the interatomic interaction, on the relative distance between atoms, $r_{ij} = |\mathbf r_i -\mathbf r_j|$.
{ The variational energies of the clusters were computed using the HFD-He interatomic potential, formulated by Aziz and collaborators \cite{azi87}. This potential is widely employed in the study of $^4$He$_N$ clusters \cite{yat22,gua06} due to its accurate modeling of the interactions between atoms in the Hamiltonian of \eq{hamil}. Exclusively to gain further insights into the optimization process, the Lennard-Jones [12-6] (LJ) potential was also utilized at this stage. The HFD-He potential offers a significantly higher level of accuracy for the system under investigation compared to the LJ potential. The minimum of the HFD-He potential is slightly shifted to large interatomic separations and is marginally deeper in comparison to the LJ interaction for helium atoms. A comparison of these potentials is presented in Fig.\ref{fig:fig01}. Among the very accurate potentials available to describe helium atom interactions, one of the motivations for choosing the HFD-He potential is the availability of existing literature results for the clusters we have analyzed.
}

The neural network architecture was built in search of efficiency. Thus, translational invariance and the symmetrical character of the wave function were encoded, since the Hamiltonian commutes with the translation operator and the wave function must obey Bose-Einstein statistics. The encoded neural network quantum state was introduced to model the ground state wave function
\begin{equation}\label{eq:psib}
\Psi_{\rm DNN}^{\rm B} (R) = 
\exp\left[ 
-\frac 1 2 \sum\limits_{i<j}^N r_{ij}^{-5}
\right]
\left(
\sum\limits_{\alpha=1}^8 \omega_\alpha \chi_\alpha(R)
\right),
\end{equation}
where $R=\{\mathbf r_i \ | \ i=1,...,N\}$ is a set of space coordinates associated with the $N$ particles that form the many-body system. The first term of this trial function is 
a product of two-body terms of the Jastrow form with a pseudopotential of the McMillan form \cite{mcm65}, which captures short-range correlations. It is multiplied by a sum of eight functions $\chi_\alpha$ weighted by variational parameters $\omega_\alpha$. This is an accurate function that includes Feynman and Cohen backflow \cite{fey56} correlations that cause an atom position to be dependent on the whole system configuration. In fact, it goes beyond the conventional manner \cite{sch81} in which backflow correlations are included. It replaces a single function by a set of functions that includes backflow.

{
The simplicity of the trial function in \eq{psib}, with only the variational parameters $\omega_\alpha$ explicitly stated, may obscure a limitation of the variational method. This limitation arises from the presence of a substantial number of parameters embedded within the functions $\chi_\alpha$,   requiring optimization, which makes this process cumbersome. In the following sections, we will delve into strategies for overcoming these challenges and demonstrate how deep neural networks can be employed to effectively represent trial functions for the study of many-body bosonic systems.}

Accurately handling atomic configurations is vital for an appropriate representation of the trial function. The information pertaining to each atom is presented as a vector
\begin{equation}\label{eq:f_input}
\mathbf f_i^0 (\mathbf r_i, \{\mathbf r_{i/}\}) =
\left( 
\mathbf q_i, q_i, 
\sum\limits_j \frac{\mathbf r_{ij}} N,
\sum\limits_j \frac{r_{ij}} N
\right) \ ,
\end{equation}
where $\{\mathbf r_{i/}\}$ is the set of all space points other than the $\mathbf r_i$ one, $\mathbf q_i = \mathbf r_i - \mathbf r_{\text{cm}}$ is the particle coordinate relative to the center of mass $\mathbf r_{\text{cm}}$ and $q_i = |\mathbf r_i - \mathbf r_{\text{cm}}|$.
The particle $\mathbf r_i$
is referenced as the main particle.

Any one of the four layers $\ell$ of the neural network has as input a single-atom stream $\mathbf h^\ell_i$ and a two-atoms stream $\mathbf h^\ell_{ij}$; for $\ell=0$, the streams are $\mathbf h^0_i=(\mathbf q_i,q_i)$ and $\mathbf h^0_{ij}=(\mathbf r_{ij},r_{ij})$. The average of both streams are employed to compute the intermediate single-atom stream vector $\mathbf f_i^\ell$ given by
\begin{equation}\label{eq:interstream}
\mathbf f^\ell_i = 
\left(
\mathbf h^\ell_i-
\sum\limits_i \frac {\mathbf h^\ell_i} N,
\sum\limits_j \frac {\mathbf h^\ell_{ij}} N
\right) \ .
\end{equation}
Each stream in layers with  $\ell > 0$ are connected through a linear operation followed by a non-linear one. Information from the previous layer is also transmitted in the form of residual connections when both streams have the same shape. These steps can be summarized as
\begin{eqnarray}
\mathbf h^{\ell+1}_i =
\tanh \left(
\mathbf V^\ell \mathbf f^\ell_i + \mathbf b^\ell
\right)
+\mathbf h^\ell_i \ ,
\\
\mathbf h^{\ell+1}_{ij} =
\tanh \left(
\mathbf W^\ell \mathbf h^\ell_{ij} + \mathbf c^\ell
\right)
+\mathbf h^\ell_{ij} \ .
\end{eqnarray}
The weights $\mathbf V^\ell$ and $\mathbf W^\ell$, as well as the biases $\mathbf b^\ell$ and $\mathbf c^\ell$ of the neural network, are variational parameters to be optimized. 
The final layer single-particle stream outputs $\mathbf h^L_i$
of each particle are reshaped 
into a matrix of elements $h_{\alpha\nu}$
used to construct the orbitals \begin{equation}\label{eq:part_state}
\phi_\alpha (\mathbf r_i, \{\mathbf r_{i/}\}) =
\sum\limits_\nu
h_{\alpha\nu}(\mathbf r_i,\{\mathbf r_{i/}\})
\exp[- a_{\alpha\nu} q_i ],
\end{equation}
where each main particle is correlated to the centre of mass to bind the system and to give the correct behavior asymptotically when $r_i \rightarrow \infty$. The decaying rates $a_{\alpha\nu}$ are also variational parameters.
The functions $\phi_\alpha$ remain invariant under exchanges of pairs of atoms that do not include the $i$-th atom. {This is due to the fact that the input features of the neural networks do not depend on the order of coordinates in $\{\mathbf r_{i/}\}$}. Finally, symmetric functions $\chi_\alpha$ are formed by the product
\begin{equation}\label{eq:chialpha}
\chi_\alpha(R) = 
\prod\limits_i 
\phi_\alpha (\mathbf r_i, \{\mathbf r_{i/}\}),
\end{equation}
where particle exchanges leave
$\chi_{\alpha}$ invariant due to the commutativity of the product, which allows rearrangement of the terms in any order.

{ 
The foundational architecture of our approach was inspired by FermiNet \cite{ferminet}, particularly in adopting a two-stream structure. However, there are several key distinctions between the two. Firstly, our input features $\mathbf f_i^0$ and intermediate stream vectors $\mathbf f_i^\ell$ are specialized for spin $\mathbf s=0$ particles. Additionally, our single-particle input $\mathbf h^0_i$ bounds the $i$-th particle coordinates to the cluster’s center of mass. This specialization arises from our treatment of the entire helium atom as a single particle, eliminating the Bohr-Oppenheimer approximation. This distinction is also evident in the intermediate stream vectors, where we employ subtraction, as outlined in \eq{interstream}, instead of concatenating the mean of single-particle vectors $\mathbf h_i^\ell$. 
Furthermore, our approach diverges significantly, starting from the $\mathbf h_i^L$ output single-particle stream onward, to address systems made from bosons. Nonetheless, it's worth noting that the linear combination of functions $\chi_\alpha$ in \eq{psib} is equivalent to the multi-determinant expansion introduced in the fermionic function \cite{pfa20}, establishing a final connection between the two Ansatz.

}

{There is currently a growing focus on extending the neural network Ansatz to periodic systems \cite{pes22,wil23}. Notably, Pescia and colleagues \cite{pes22} have investigated $^4$He systems in one and two dimensions. Although both \eq{f_input} and their trial function  
ensure that the input features remain invariant with respect to the symmetries of the problem, the approaches are distinct. In their work, periodic boundary conditions are implemented, while for the clusters, global translation invariance is considered. Additionally, the pooling operation, which confers permutation invariance, resembles the procedure outlined in \eq{part_state}. It's worth noting that their wave function construction is independent of system size, which proves advantageous once the network is trained. In contrast, our trial function is size-dependent, necessitating multiple training sessions that, nevertheless, also afford greater flexibility for the neural network to learn.}

Unsupervised training of the neural network depends on the sampling of the probability distribution
\begin{equation}
p(R) = 
\frac 
{ |\Psi^{\rm B}_{\rm DNN} (R)|^2 }
{ \int dR \ |\Psi^{\rm B}_{\rm DNN} (R)|^2 } \ .
\end{equation}
The expectation value of the Hamiltonian $E$ is estimated by approximating the integral $\int dR\, p(R) E_L(R)$ by the
average values of the local energy $E_L(R)=\mathcal{H}\Psi^{\rm B}_{\rm DNN}(R)/\Psi^{\rm B}_{\rm DNN}(R)$
over the sampled configurations \cite{cep79}.

The set $\theta$ of all variational parameters is optimized using a second-order gradient-based method that accounts for correlations between pairs of variational parameters through derivatives of energy $E$ and density probability $p$ \cite{pfa20}
\begin{align}
&\bm\nabla_\theta E =
2 \int dR \ 
p(R) \times \\ \nonumber
& \times \left[
\frac 
{\mathcal H \psi (R)}
{\psi (R)} -
\int dR' \ 
p(R')
\frac 
{\mathcal H \psi (R')}
{\psi (R')}
\right]
\bm\nabla_\theta
\log|\psi (R)| \ .
\end{align}
The optimization process updates the parameters repeatedly 
according to $\mathcal F^{-1} \bm\nabla_\theta E$, where $\mathcal F^{-1}$ is the inverse of the Fisher information 
matrix,
\begin{equation}
\mathcal F_{ij} = 
\int dR \ 
\frac
{\partial \log p(R)} 
{\partial \theta_i}
\frac
{\partial \log p(R)} 
{\partial \theta_j},
\end{equation}
\noindent
 which correlates pairs of parameters.

{
The extensive parameter space inherent in neural networks can be handled in different ways. Particularly, the update of neural network parameters is a crucial step in the deep learning process. The iterative solver framework \cite{vic22} is a powerful approach that can incorporate different methods to facilitate efficient parameter updates. In this work, we focus on employing one of these methods that may potentially be utilized within the iterative solver framework, namely the Kronecker-factored approximate curvature (K-FAC) method \cite{mar15}. This method hinges on two critical principles. Firstly, it approximates the Fisher information matrix as block diagonal. This insight, rooted in the domain of neural networks, recognizes that for functions represented by neural networks, elements $F_{ij}$ of the Fisher matrix effectively become zero if the network parameters $\Theta_i$ and $\Theta_j$ pertain to different layers. This approximation resonates with the hierarchical structure of neural networks, where parameters across distinct layers tend to exhibit a high degree of independence.}

{Secondly, to augment computational efficiency, the method employs Kronecker factorization to approximate the inverse of each of these blocks, including slight modifications to extend its applicability to unnormalized probability densities \cite{pfa20}. This technique offers a highly effective approach for managing the large-scale matrices inherent in neural network-based functions, thereby rendering the optimization process computationally manageable.}

{Collectively, these considerations underpin the methodology, facilitating the optimization of trial functions based on neural networks for intricate systems. In our specific case, this approach leads to a significant reduction in computational complexity, replacing the inversion of matrices with dimensions on the order of $10^5$ with the inversion of ten matrices, each with dimensions of about $10^2$.
}

{In the optimization process, the selection of hyperparameters plays a crucial role in ensuring stability and efficiency. Many of these parameters can be extracted from the FermiNet approach \cite{ferminet}. The typical parameters for the simulations are provided in \tab{optim_hyperpar}. The step size of proposed movements is not included, since it is dynamically adjusted during the simulations. It is worth noting that the width of the layers can be reduced for smaller clusters, particularly for the single-atom stream width.}

{\begin{table}[!ht]
\caption{{Typical hyperparameter values employed in the simulations. An open-source version of the code used in this work  %can be found at \url{https://github.com/freitas-esw/bosenet-helium-clusters}
to optimize the trial functions is available \cite{bnnhc}.}
}
\centering
{\begin{tabular}{ll}\hline\hline
%Parameter &  \\ \hline
Batch size & 8192 \\
Single-atom channel width & 128 \\
Two-atom channel width    & 16 \\ 
Symmetric functions & 8 \\
Number of layers & 4 \\
Learning rate & 0.001 \\ 
Learning rate decay & 1.0 \\
Learning rate delay & 10000 \\
Training iterations & $2 \times 10^5$ \\
Damping & 0.001 \\
Momentum & 0.0 \\
Covariance moving decay & 0.95\\
Norm constrain & 0.001 \\
\hline\hline
\end{tabular}}
\label{tab:optim_hyperpar}
\end{table}
}

{It is worthwhile to discuss why the results were obtained without a parameterization of the Jastrow factor in \eq{psib}, as is typically done in variational calculations for systems composed of helium atoms. The main motivation for this choice was to enhance the efficiency of the optimization process. Preliminary tests, where this factor was parameterized, showed an unstable optimization process due to the intense character of the helium short-range interatomic interaction. While the HFD-He potential does not diverge at the origin, it does exhibit considerably high values for small pair separations. In finite-time runs, no sampled configuration will feature a pair at distances smaller than approximately 1 \AA. However, eventually sampled configurations may involve a pair separation where the potential still has a substantial energy, and a local energy that significantly deviates from the average value. In principle, the Laplacian of the wave function should perfectly cancel out this value. However, the proposed trial wave function is only an approximation for the description of the ground state, and this perfect cancellation does not occur. Consequently, these imprecise estimations can significantly impede the optimization process by pushing variational parameters away from their optimal values and rendering the optimization process excessively slow.}

{
The absence of the parameterization of the Jastrow factor in a trial function represented by a neural network for a system composed of helium atoms stands in contrast to approaches used in systems governed by Coulombic interactions \cite{pfa20,gao22,von23}. The latter architectures prove especially effective in handling electron-electron correlations, scenarios where the Coulomb potential exhibits relatively predictable and manageable behavior. }

\begin{figure}
    \centering
    \includegraphics[width=.4\textwidth]{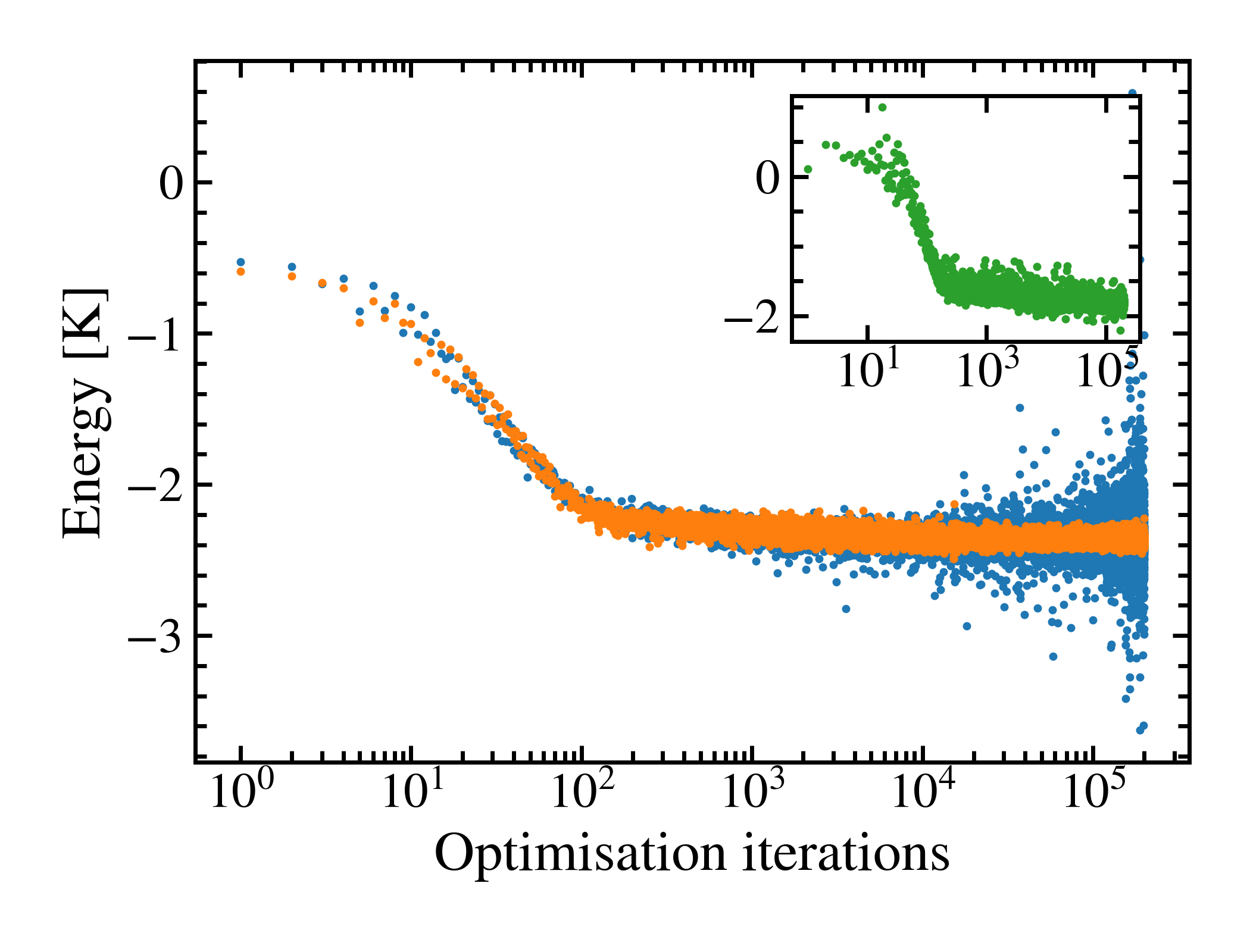}
    \caption{The total energy, without associated uncertainties, is presented to elucidate its evolution as a function of the optimization step for a cluster of $N=6$ atoms employing the HFD-He potential (depicted as blue dots). The influence of incorporating a hard sphere (HS) potential is represented by the orange dots (refer to the text for details). The inset illustrates the optimization process for the Lennard-Jones system.}
    \label{fig:fig1}
\end{figure}

{These findings were validated through a simulation employing a Lennard-Jones potential with parameters $\varepsilon=10.22$ K and $\sigma=2.556$ Å for a 6-atom cluster. In this scenario, the Jastrow factor in \eq{psib}, incorporating a McMillan pseudopotential with an $r^{-5}$ dependency, ensures an asymptotically correct solution to the Schrödinger equation as $r \rightarrow 0$. The performance of the neural network with this Jastrow factor choice is depicted in the inset of \fig{fig1}, illustrating the total energies throughout the optimization process as a function of the performed iterations. The uncertainties associated with the energy are omitted due to significant variances in the estimated values during the initial iterations, making it difficult to observe the energy's progression as the optimization advances. Following the optimization, a standard VMC simulation with fixed variational parameters yielded a total binding energy of -1.8007(1) K, a value notably higher than that obtained with the HFD-He potential. This is the only result presented for the LJ potential, since its inclusion in this work was solely for the purpose of investigating the optimization stability.}

{The optimization of the parameters of the DNN Ansatz, \eq{psib}, is performed to minimize the energy. Initially, the learning process was focused on determining parameters that yielded reasonable energy values. Subsequently, without an explicit directive, the neural network learned to reduce the variance, leading to more precise results as a consequence. This behavior can be seen  in \fig{fig1}.}
More specifically, each optimization iteration consisted in obtaining a set of sampled configurations by applying the Metropolis algorithm followed by computing the energy and its gradients, the Fisher matrix, and updating the parameters to conclude the iteration. Typically, the number of configurations, \textit{i.e.} the batch size, in each step was $2^{13}$. 
After reaching a converged energy with a small variance, the optimization process was stopped. Then, a standard variational simulation was  performed to get the final results and their associated variances.

{At the beginning of the optimization process, large fluctuations in the total energy of the system often obscure the evolution of this quantity. To provide a clearer view of its progression during the optimization process, the uncertainties of the values shown in \fig{fig1} are not displayed, as in the case of the LJ system.}

{Moreover, during the optimization, it was observed that this process evolved in a less than optimal manner, exhibiting slight instabilities after obtaining parameters that reasonably described the cluster energy. To address this challenge, we incorporated a hard sphere (HS) potential into the interatomic interaction given by the HFD-He potential during the course of the optimization.}
This addition addresses the issue that the $r^{-5}$ dependence in the Jastrow factor of $\Psi_{\rm DNN}^{\rm B}$, \eq{psib}, does not properly account for, in terms of behavior at short pair distances. The HS potential prevents atoms from getting too close to each other, overcoming the limitations of the Jastrow factor at small distances, which can lead to low-probability configurations.

The specific chosen value  (1.8 \AA) of the HS radius is not critical and did not produce any significant changes in the energy estimate. In fact, the weak unitarity satisfied by the helium clusters, makes the addition of an HS potential irrelevant, which will be discussed later. The optimization process with an infinite barrier at the HS radius is also shown in \fig{fig1}. The results indicate that the inadequate functional dependence of the pseudopotential in the Jastrow factor is a source of instability during the optimization process. The inclusion of the HS potential was able to lower the variances. Therefore, for all optimizations performed (and only in this stage), the HS potential was added to the HFD-He interatomic interaction.

\begin{table}[ht]
\caption{%
Kinetic $\langle \mathcal T \rangle$ and total ground state $\langle E \rangle$ energies obtained with the DNN ansatz in units of Kelvin are presented for a $^4$He cluster of size $N$. The fourth column shows DMC results from the literature \cite{gua06}.}
\label{tab:et}
\centering
\begin{tabular}{ccccc}
$N$ &
$\langle \mathcal T \rangle$&
$\langle E \rangle$&
DMC   \\[2mm]
\hline\\[-2mm]
%\colrule
2  & 0.1246(4) & -0.002142(6) & - \\
3  & 1.695(2)  & -0.13323(9) & -0.135(2)  \\
4  & 4.353(3)  & -0.5775(1)  & -0.573(2)  \\
5  & 7.745(4)  & -1.3341(2)  & -1.334(2)  \\
6  & 11.666(4) & -2.3710(3)  & -2.367(3) \\
7  & 16.145(5) & -3.6510(4)  & -3.646(4)  \\
8  & 20.997(6) & -5.1448(4)  & -5.144(5)  \\
9  & 26.212(7) & -6.8207(9)  & -6.827(6)  \\
10 & 31.868(8) & -8.6766(7)  & -8.673(6)  \\
11 & 37.72(1)  & -10.665(2)  & - \\
12 & 43.50(1)  & -12.801(2)  & - \\
13 & 48.79(1)  & -15.069(2)  & - \\
14 & 55.92(1)  & -17.444(2)  & - \\
\hline
\end{tabular}
\end{table}

\section*{Results}

The reported final cluster energies were obtained in a standard VMC calculation with the interatomic interaction $V$ given exclusively by the HFD-He potential while keeping the $\Psi_{\rm DNN}^{\rm B}$ parameters fixed. These results are presented in \tab{et}.
Comparisons of the variational results with those from the DMC show that within statistical uncertainty they are in excellent agreement. This is remarkable because the DMC results are in principle exact in the Monte Carlo sense.  For most of the clusters, the variational results are within one statistical standard deviation from the DMC results.
\begin{figure}
    \centering
\includegraphics[width=.4\textwidth]{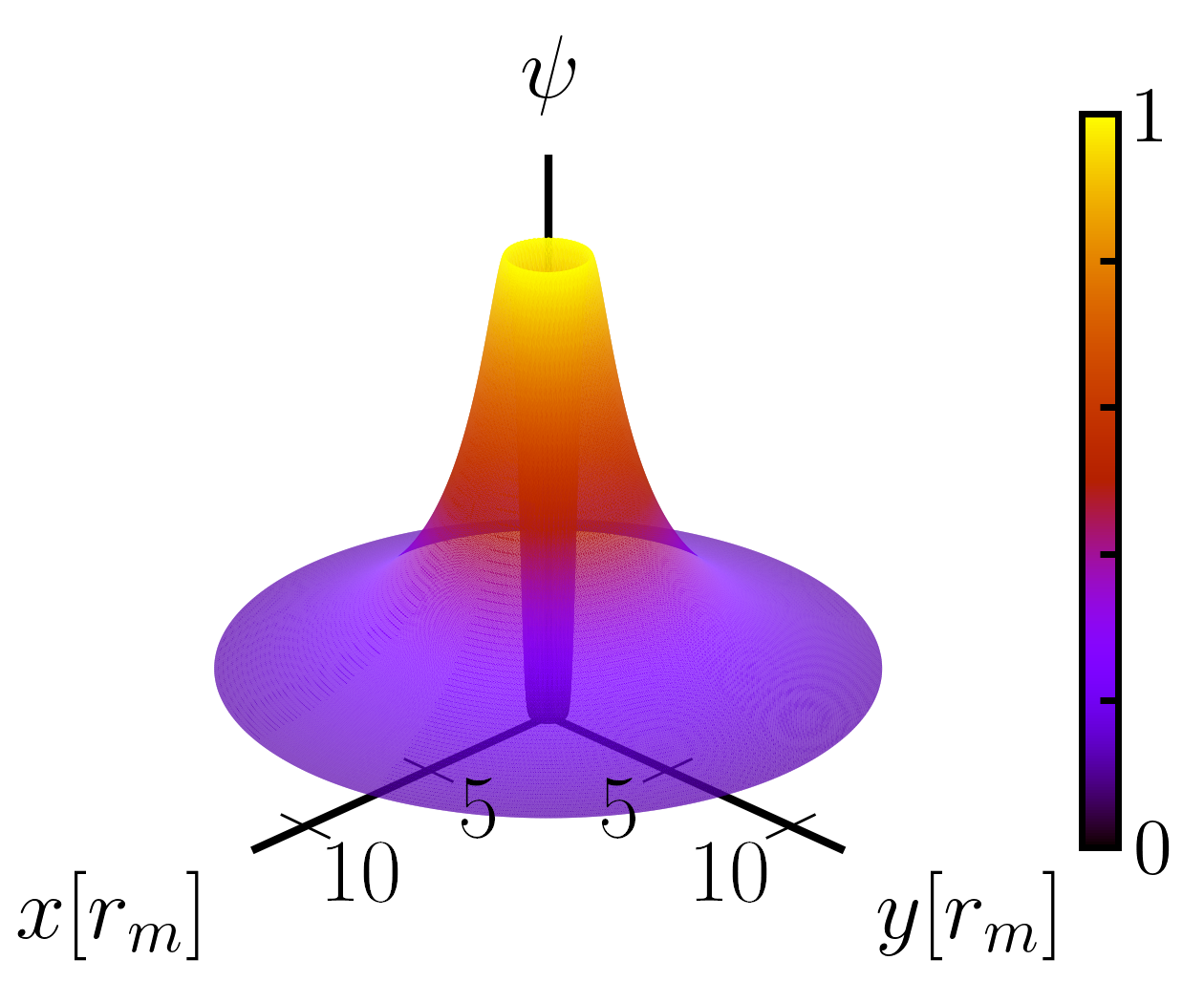}
    \caption{%
    The wave function of the helium dimer is shown, with light surface colors emphasizing regions of large probability amplitude in an arbitrary scale. The space coordinates are given in terms of $r_m=2.963$nm, where the potential is minimum. }
    \label{fig:n2wfn}
\end{figure}

A three-dimensional plot of the two-body wave function is displayed in \fig{n2wfn}, where one of the atoms is at the origin and the probability amplitudes are shown for points $(x,y)$ in the plane $z=0$. 
{The probability amplitude exhibits a peak at an interatomic separation of approximately 3.8 \AA. The bond length, determined by the average value of the pair distance $\langle r_{ij}\rangle$, is $36.6 \pm 0.6$ \AA.}
However, small pair separation is not unexpected, since the interatomic interaction used does not include relativistic and quantum electrodynamics contributions \cite{prz10}. The $^4$He$_2$ dimer is known to exist mostly in the quantum tunneling regime in a quantum halo state  \cite{zel16etal}. The plot demonstrates that even though radial symmetry was not explicitly imposed, the neural network was able to learn it.

\begin{figure}
    \centering
\includegraphics[width=.4\textwidth]{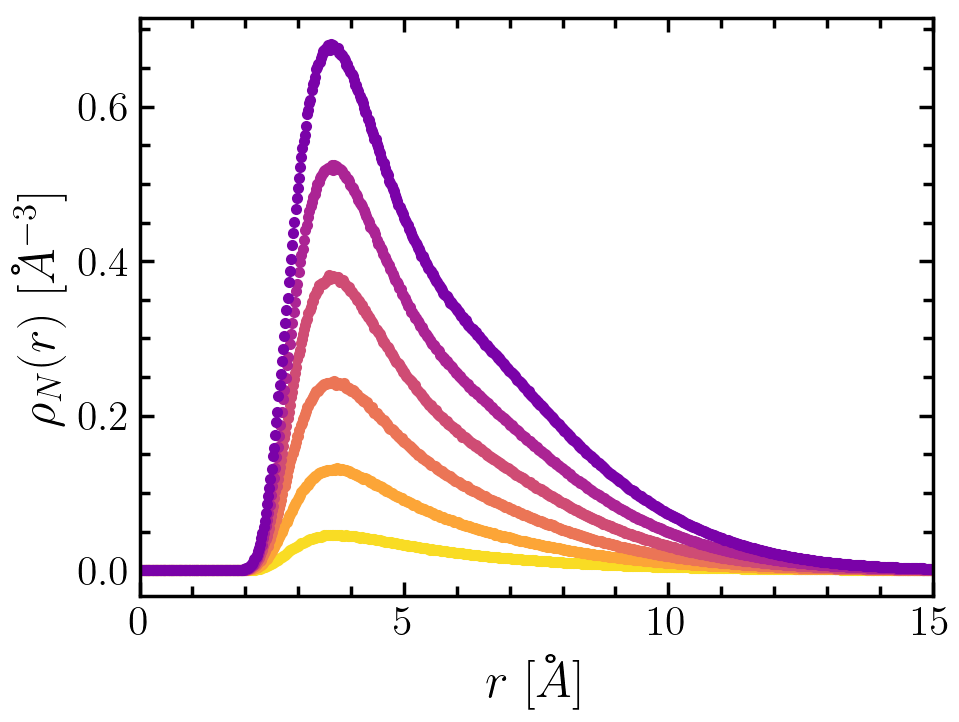}
    \caption{Pair density function for $^4$He$_{N}$ clusters with $N=\{4,6,8,12,14\}$ atoms displayed by colors from light to dark.}
    \label{fig:fig2}
\end{figure}

The pair density function $\rho_N(r)$ for a cluster of $N$ atoms is estimated by the operator
\begin{align}
    \hat \rho_N(r) 
    = 
    \sum_{i<j}^N
    \frac{\delta(r_{ij} - r)}{r^2}.
\end{align}
The integration of $\rho_N(r)$ in the whole space gives
$\int \rho_N(r) r^2 \mathrm{d} r = N(N-1)/2$. 
The pair density function is shown in \fig{fig2}.

{Helium clusters are bound by the attractive van der Waals interaction, typically characterized by a $(-C_6/r^6)$ tail. This phenomenon arises due to the zero-point fluctuations of atomic dipole moments, leading to short-range correlations. In the study of these clusters, attention has recently been directed to the universal character of the short-range correlations \cite{yat22,baz20}  by adapting the Tan relations  \cite{tan08a,tan08b,tan08c} to clusters \cite{mil18,wer12pra,wer12ferm}. Within the framework of weak unitarity, the concept of contact that emerges from these relations serves to quantify the density of closely interacting pairs in strongly interacting systems, offering  insights into their behavior at short distances.

While strong universality is not expected in helium clusters due to their characteristic distances (with an average pair distance of approximately 5 \AA\ and a van der Waals length of about 5.4 \AA\ \cite{baz20}), it is well established \cite{mcm65} that the behavior of helium systems is primarily influenced by the closest pair of atoms.  Nonetheless, provided there exists a strongly interacting model at short distances, the $N$-body wave function can still be factorized into the product of a universal two-body function and a state-dependent function \cite{baz20}. This remains valid, because in close proximity, a correlated pair is minimally influenced by the surrounding particles. Consequently, its two-body wave function should exhibit consistency regardless of the system's size or state. This scenario  is called weak universality \cite{baz20}.}

%Helium clusters are bound by an attractive van der Waals interaction tail $(-C_6/r^6)$, due to the zero point fluctuations of atomic dipole moments, leading to short-range correlations. In the study of these clusters, attention has recently been directed to the universal character of the short-range correlations \cite{yat22,baz20} by adapting the Tan relations \cite{tan08a,tan08b,tan08c}  to clusters \cite{mil18,wer12pra,wer12ferm}. However, a strong universality is not expected, because the average pair distance for clusters with $N>3$ is 5 \AA\ and the characteristic potential range given by the van der Waals lengthis approximately 5.4 \AA\ \cite{baz20}. Nevertheless, for the helium systems, the overall behavior is, in general,  determined by the closest pair of atoms, so that the wave function can be thought of as a factorized universal two-body term times a state-dependent term. This situation is known as weak universality \cite{baz20}. 

The $r$-independent two-body contact $C_2^{(N)}$ can be estimated through the pair density function
\begin{align}    {\rho_N(r)}_{\large\ \overrightarrow{\text{small }r}\ } C^{(N)}_2 \rho_2(r),
\end{align}
by definition $C_2^{(2)}=1$. The $C_2^{(N)}$ are treated as a fit parameter adjusted for values of $r$ up to the maximum value of $\rho_N(r)$ in the minimization of the integral $\int (\rho_N(r) - C_2^{(N)} \rho_2(r))^2 {\rm d}r$.

\begin{figure}
    \centering
\includegraphics[width=.4\textwidth]{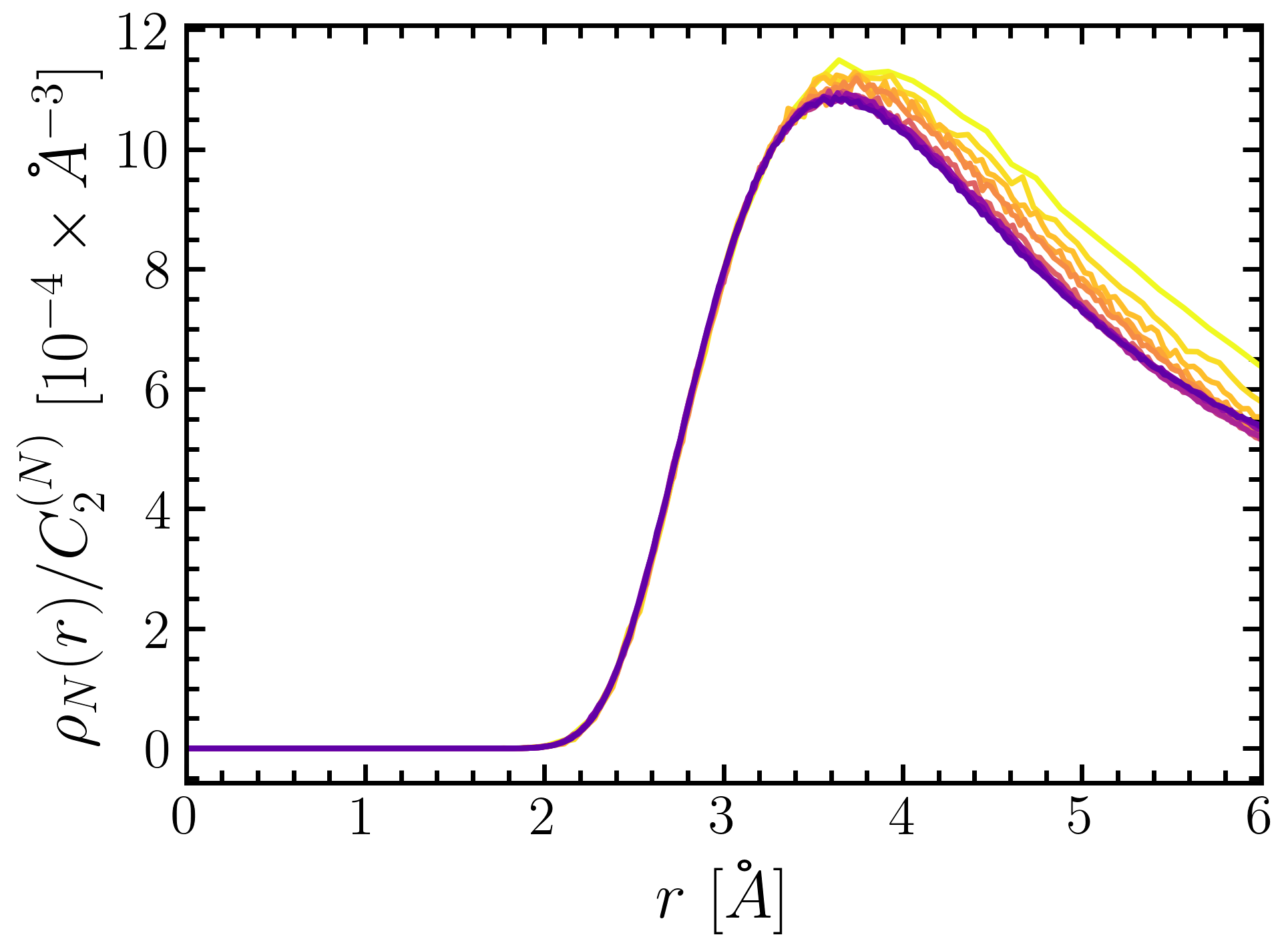}
    \caption{Pair density functions of $^4$He$_N$ clusters with $N=2$ to $N=14$ atoms normalized with the appropriate contact $C_2^{(N)}$ shown by colors from light to dark.}
    \label{fig:3}
\end{figure}

The results of the pair density functions for $^4$He$_N$ clusters, ranging from $N = 2$ to $N = 14$ atoms, normalized by contacts $C_2^{(N)}$ as a function of the radial distance, are displayed in \fig{3}. As shown, there is a clear collapse of the pair density functions for all cluster sizes. These results indicate that even the largest clusters, with up to 14 atoms, exhibit a weak universality. This is consistent with the coalescence of N-body atoms in a Bosonic system, as previously predicted \cite{wer12pra}. The pair-atom contact $C_{2+1}^{(N)}$ \cite{yat22} will be discussed in a separate publication.

\section*{Final comments}

The use of neural network-based representation of the trial function for a bosonic system yielded results for core properties of $^4$He$_N$ clusters that are in agreement with those obtained using the exact diffusion Monte Carlo method.

The implemented approach for the deep neural network does not rely on external information. The chosen representation of the trial function, by explicitly satisfying the Bose-Einstein statistics, avoids machine time spent considering non-physical solutions. Additionally, by satisfying global translation symmetry, the representation eliminates the need to consider infinite degenerate coordinates that represent the same physical situation. These features of the representation allow for an efficient optimization process, able to obtain results with modest computational resources.

Our results also demonstrated the importance,  in the optimization process, of two-body correlation factors with the appropriate properties at the asymptotic limit $r \rightarrow 0$. The weak universality of $^4$He$_N$ clusters enabled the inclusion of an HS potential in the atomic interaction, which stabilized the optimization process by eliminating configurations with very low probability, where two atoms are too close together. Previous literature \cite{yat22,lut17,vit92} has presented various proposals for the two-body correlation factors of $^4$He$_N$ clusters. The optimization process, due to its features at the limit $r \rightarrow 0$, could be valuable in determining the correlation factor that most accurately describes the system.
An improved functional form of the two-body correlation factor in $\Psi_{\rm DNN}^{\rm B}$ may remove the need for a cut-off in the interatomic interaction during optimization. An alternative and simpler approach that could also improve optimization is to use a soft sphere potential instead of an HS potential, which would allow the wave function to smoothly approach zero at short distance separations between atoms.

The weak universality of $^4$He$_N$ clusters is shown through the collapse, up to the maxima, of the pair density functions normalized by their respective contacts $C^{(N)}_2$.
These results were obtained by using a deep neural network to represent the trial function $\Psi_{\rm DNN}^{\rm B}$. This approach is validated both by previous findings \cite{yat22,baz20} and the specific analysis performed in this work.

One way to expand on this research is to study larger clusters and use the liquid drop model to extract the contact $C^{(2)}_2$ for bulk liquid $^4$He. Another avenue for investigation could be to look into the effects of impurities in the clusters.

\begin{acknowledgments}
   WF is thankful for useful discussions with Dr. Markus Holzmann and Dr. Matthew Foulkes. Simulations were performed in part at the Centro Nacional de Processamento de Alto Desempenho em S\~ao Paulo (CENAPAD-SP). The authors acknowledge financial support from the Brazilian agency, Funda\c{c}\~ao de Amparo \`a Pesquisa do Estado de S\~ao Paulo (FAPESP), projects Proc. No. 2016/17612-7 and Proc. No. 2020/10505-6.
\end{acknowledgments}

\bibliographystyle{unsrtnat}
\bibliography{sav} % Produces the bibliography via BibTeX.

\end{document}